# Anti-Hong-Ou-Mandel interference by coherent perfect absorption of entangled photons

Anton N. Vetlugin[1*], Ruixiang Guo[1], Cesare Soci[1*], and Nikolay I. Zheludev[1,2]

[1]*Centre for Disruptive Photonic Technologies, SPMS, TPI, Nanyang Technological University, Singapore 637371*
[2]*Optoelectronics Research Centre and Centre for Photonic Metamaterials, University of Southampton, Southampton SO17 1BJ, United Kingdom*
*Corresponding author. Email: a.vetlugin@ntu.edu.sg (A.N.V.); csoci@ntu.edu.sg (C.S.)

Two-photon interference, known as the Hong-Ou-Mandel effect, has colossal implications for quantum technology. It was observed in 1987 with two photodetectors monitoring outputs of the beamsplitter illuminated by photon pairs: the coincidence rate of the detectors drops to zero when detected photons overlap in time. More broadly, bosons (e.g., photons) coalesce while fermions (e.g., electrons) anti-coalesce when interfering on a lossless beamsplitter. Quantum interference of bosons and fermions can be tested in a single – photonics platform, where bosonic and fermionic states are artificially created as pairs of entangled photons with symmetric and anti-symmetric spatial wavefunctions. We observed that interference on a lossy beamsplitter, or a subwavelength coherent absorber reverses quantum interference in such a way that bosonic states anti-coalesce while fermionic states exhibit coalescent-like behavior. The ability to generate states of light with different statistics and manipulate their interference offers important opportunities for quantum information and metrology.

*Introduction*.-The Hong-Ou-Mandel (HOM) effect [1] occurs when two indistinguishable particles interfere on a beamsplitter. In such a process, the output state depends on the symmetry of the two-particle wavefunction. “Bosons coalesce and fermions anti-coalesce” is a common resume of the HOM interference, Fig. 1A, which was supported by numerous experiments [2-11]. Besides being one of the most striking fundamental phenomena, the HOM effect plays a key role in various applications, from timing measurement [12] and lithography [13] to quantum computation [14-16], boson sampling [16,17], and communication [18,19].

The nature of the beamsplitter plays an essential role in the HOM effect [20]. It is commonly accepted that the presence of dissipative channels of a beamsplitter hampers the HOM interference; thus, losses are usually averted in experiments. However, the presence of dissipation may induce non-trivial effects. For instance, theory predicts that a carefully designed lossy beamsplitter may reverse the outcome of two-boson interference, from coalescence to anti-coalescence [21], Fig. 1B. This prediction was experimentally verified in a plasmonic system [22]. Similarly, theory predicts that fermions interfering on a lossy beamsplitter may exhibit coalescent-like behavior [22], Fig. 1B.

In quantum optics, entangled photons may possess either symmetric (“bosonic”) or anti-symmetric (“fermionic”) *spatial* wavefunctions depending on the polarization Bell state [23-25]. Different regimes of interference of the symmetric and anti-symmetric wavefunctions are exploited in quantum information and metrology applications where propagating entangled photons are dynamically converted into different states [26-35]. The ability to switch between the “fermionic” behavior of bosonic spatial wavefunction and the “bosonic” behavior of fermionic spatial wavefunction may therefore enrich these protocols and provide new approaches for quantum light manipulation. Recently, this reversed effect was demonstrated for symmetric wavefunctions of (non-entangled) photon pairs by exploiting a grating metasurface as a lossy beamsplitter [36]. Nevertheless, the lack of entanglement in the system and polarization selectivity of the metasurface make this approach unsuitable for the non-unitary processing of anti-symmetric spatial wavefunctions.

Here, we experimentally study the interference of symmetric and anti-symmetric spatial wavefunctions of entangled photons in the absence and presence of dissipation. In the latter case, we exploit the coherent perfect absorption (CPA) scheme where the counter-propagating photons interfere on thin absorptive layers. Albeit dissipative, the coherent perfect absorption scheme preserves light coherence [37]. While the symmetric (anti-symmetric) spatial wavefunction coalesces (anti-coalesces) as a result of dissipation-free interference, dissipative interference by CPA yields the opposite outcome, with the symmetric (anti-symmetric) spatial wavefunction showing anti-coalescent (coalescent-like) behavior. To demonstrate this reversed – 'anti-Hong-Ou-Mandel' interference, we follow the transformation method of CPA of quantum light [37] by mapping the two-photon wavefunction from the original travelling wave basis into the basis of standing waves, performing non-unitary transformation in this basis and, finally, mapping the wavefunctions back into the travelling wave basis (see details below). The non-unitary transformation is implemented by a subwavelength absorber (lossy beamsplitter) designed to be sensitive to the standing wave symmetry. Our approach is polarization insensitive, thus allowing us to perform the non-unitary transformation of any two-photon states. This opens access to the processing of entangled photon states, including states with fermionic symmetry, which has not been demonstrated before.

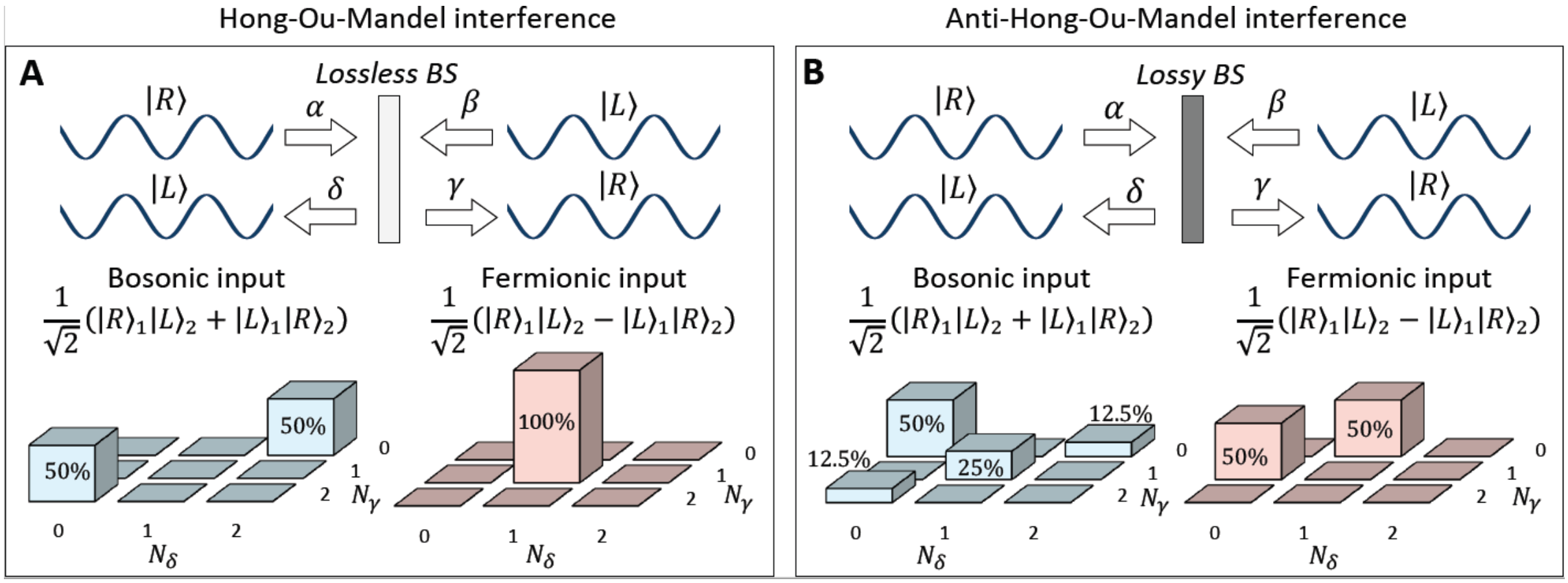

FIG. 1. **From Hong-Ou-Mandel to anti-Hong-Ou-Mandel interference**. A: In the HOM interference, particles coalesce (anti-coalesce) if input two-photon spatial wavefunction possess bosonic (fermionic) symmetry. B: In the anti-HOM interference, particles show reversed behavior: bosonic (fermionic) spatial wavefunction exhibits anti-coalescent (coalescent-like) behavior.

Bosonic and fermionic anti-HOM interference, along with other non-unitary phenomena like entanglement generation by dissipation [38,39] and coherent perfect absorption [40-42] of quantum light [43-52], emphasizes the fundamental and practical importance of coherence preserving dissipation in quantum optics.

*Hong-Ou-Mandel and anti-Hong-Ou-Mandel interference*.- We consider two indistinguishable photons in the counter-propagating geometry, Fig. 1. Photons are bosons that restrict the symmetry of the *total* wavefunction under the permutation of particles. The total wavefunction, $|\Psi_{\mathrm{pol}}\rangle\otimes|\Psi_{\mathrm{sp}}\rangle$, contains polarization $|\Psi_{\mathrm{pol}}\rangle$ and spatial $|\Psi_{\mathrm{sp}}\rangle$ components. For instance, symmetric polarization Bell state, $|\Psi^{(s)}_{\mathrm{pol}}\rangle=\frac{1}{\sqrt{2}}(|H\rangle_1|V\rangle_2+|V\rangle_1|H\rangle_2)$, is accompanied by symmetric (bosonic) spatial wavefunction [53],

$$\left|\Psi^{(\mathrm{s})}_{\mathrm{sp}}\right\rangle=\frac{1}{\sqrt{2}}(|L\rangle_1|R\rangle_2+|R\rangle_1|L\rangle_2), \tag{1}$$

while anti-symmetric polarization Bell state, $|\Psi^{(as)}_{\mathrm{pol}}\rangle=\frac{1}{\sqrt{2}}(|H\rangle_1|V\rangle_2-|V\rangle_1|H\rangle_2)$, is accompanied by anti-symmetric (fermionic) spatial wavefunction,

$$\left|\Psi^{(\mathrm{as})}_{\mathrm{sp}}\right\rangle=\frac{1}{\sqrt{2}}(|L\rangle_1|R\rangle_2-|R\rangle_1|L\rangle_2). \tag{2}$$

Here $|H\rangle$ ($|V\rangle$) defines the horizontal (vertical) polarization and $|L\rangle$ ($|R\rangle$) defines the left (right) direction of propagation. The lower index identifies the photon's number. First, we consider the case when photons enter a *lossless* beamsplitter (through input ports $\alpha$ and $\beta$, Fig. 1A) characterized by amplitude transmission $t$ and reflection $r$ coefficients and for which we assume [54]

$$t = |t| \text{ and } r = \pm i|r|. \tag{3}$$

The beamsplitter redistributes each photon between two output ports as $|L\rangle \to t|L\rangle + r|R\rangle$ and $|R\rangle \to r|L\rangle + t|R\rangle$. Photons with the symmetric spatial wavefunction tend to leave the beamsplitter through the same output port [2], and they do so always if $|t| = |r| = 1/\sqrt{2}$: $\left|\Psi_{\text{sp}}^{(\text{s})}\right\rangle \to \frac{1}{\sqrt{2}}(|L\rangle_1|L\rangle_2 + |R\rangle_1|R\rangle_2)$. In this case, the output photon number distribution contains only the terms $N_\gamma = 2/N_\delta = 0$ and $N_\gamma = 0/N_\delta = 2$ with a lack of $N_\gamma = 1/N_\delta = 1$ contribution (left blue histogram in Fig. 1A). Here, $N_\gamma$ ($N_\delta$) is the average number of photons presented at the beamsplitter's output port $\gamma$ ($\delta$). In contrast, the anti-symmetric spatial wavefunction is "immune" to any lossless beamsplitter transformation, $\left|\Psi_{\text{sp}}^{(\text{as})}\right\rangle \to \left|\Psi_{\text{sp}}^{(\text{as})}\right\rangle$, and photons are always found in different output ports with characteristic 'anti-coalescent' term $N_\gamma = 1/N_\delta = 1$ in the output distribution (right red histogram in Fig. 1A) [26,54]. This transformation of symmetric and anti-symmetric spatial wavefunctions by the lossless beamsplitter is known as the HOM effect. We note that alternatively, a transition from coalescence to anti-coalescence of photon pairs can be observed by exploiting other degrees of freedom, such as time bins [55], or by interfering two-photon states at the detectors, not at the beamsplitter [56,57].

The magnitude of the 'anti-coalescent' component ($N_\gamma = 1/N_\delta = 1$) of the bosonic output distribution, Fig. 1A, is defined by the coefficient $P_{11} = |t^2 + r^2|^2$ where amplitudes of two possible outcomes – both photons are transmitted or both photons are reflected, are summed up [21]. The coefficient $P_{11}$ depends on both the absolute values of $t$ and $r$ and their mutual phase. For a lossless beamsplitter, the phase relation (3) between $t$ and $r$ dictates suppression of the 'anti-coalescent' component. To alter the phase relation between $t$ and $r$, the beamsplitter should possess dissipation channels. For in-phase or out-of-phase $t$ and $r$, terms $t^2$ and $r^2$ in $P_{11}$ summed up coherently, increasing the magnitude of the 'anti-coalescent' component, thus completely changing the interference patterns. In the extreme case of the lossy beamsplitter with

$$t = \pm r = 0.5, \tag{4}$$

photon pairs with the symmetric spatial wavefunction (1) experience *a probabilistic two-photon absorption* [21,37] with a high probability of the 'anti-coalescent' $N_\gamma = 1/N_\delta = 1$ output (left blue distribution in Fig. 1B). Conversely, photons with the anti-symmetric spatial wavefunction (2) experience *a deterministic one-photon absorption* [22,37] and the 'anti-coalescent' term $N_\gamma = 1/N_\delta = 1$ vanishes (right red distribution in Fig. 1B). As a result, dissipation channels of the beamsplitter induce the anti-HOM interference patterns: photons with the symmetric (anti-symmetric) spatial wavefunction show anti-coalescent (coalescent-like) behavior.

To demonstrate the anti-HOM interference with entangled photons, a beamsplitter should have an equal optical response defined in (4) at any polarization. For instance, plasmonic metasurfaces [45,46,58,59] may be used where polarization-independent operation [47] could be achieved by the meta-molecules design optimization. Here, instead of pursuing the optical response (4), we develop an alternative approach oriented toward reconstructing the output photon distribution. Our approach is based on the coherent perfect absorption of quantum light – the non-unitary transformation of standing waves by a thin absorber [37], which allows us to achieve the two-photon absorption of the symmetric spatial wavefunction (1) and the one-photon absorption of the anti-symmetric spatial wavefunction (2) and, thus, demonstrate the anti-HOM interference.

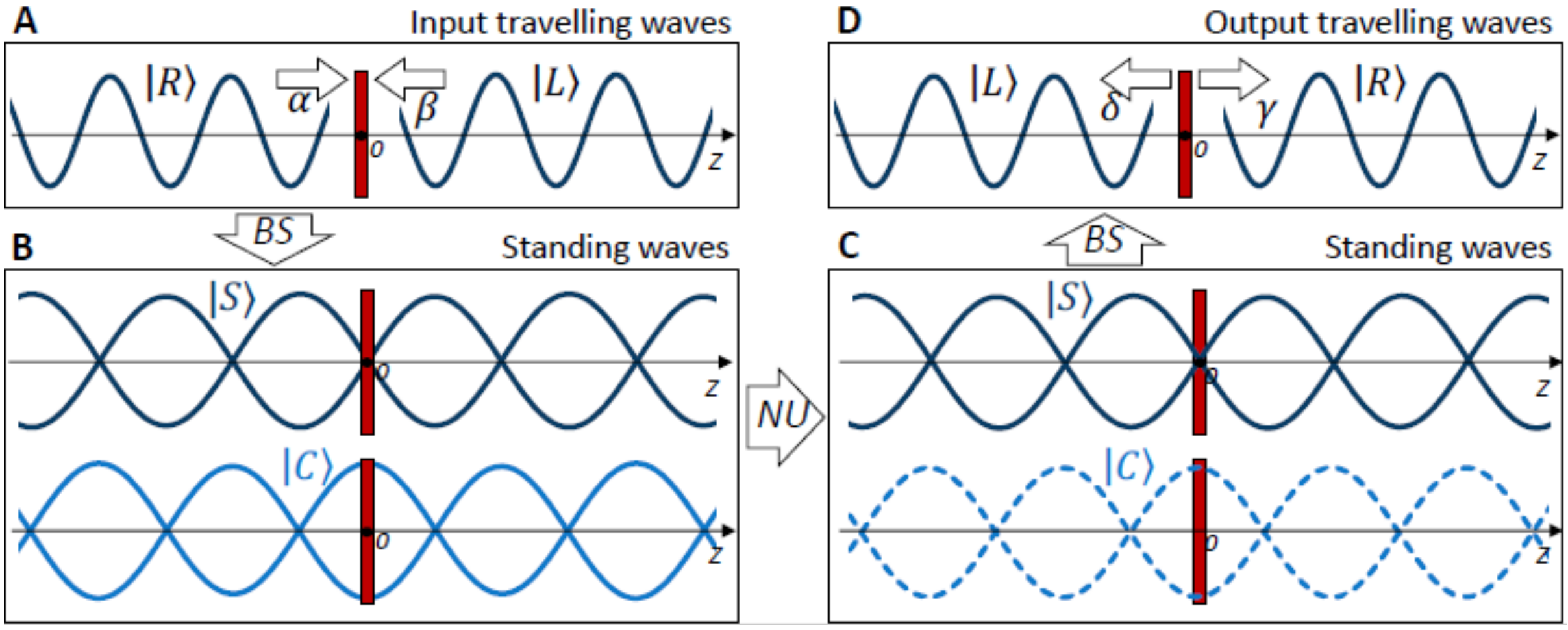

FIG. 2. **Anti-HOM interference by coherent perfect absorption of photon pairs.** Two photons occupying travelling waves (A) excite two – cosine and sine – standing waves (B). Coherent absorber (lossy beamsplitter) absorbs all excitations of the cosine wave not affecting the sine wave (C) (NU – non-unitary transformation). Survived excitations of the sine wave feeds up the outgoing travelling waves (D).

Two counter-propagating travelling waves, carrying a single photon each, Fig. 2A, $\sim e^{i(kz-\omega\tau)}$ and $\sim e^{i(-kz-\omega\tau)}$, excite two standing waves with $\cos(kz)$ and $\sin(kz)$ spatial oscillations ($k = 2\pi/\lambda$ is the wave number, $\lambda$ is the central wavelength, $\omega = ck$ is the angular frequency, and $c$ is the speed of light), Fig. 2B. This process is described by a beamsplitter-like transformation, and photons are distributed between the standing waves as [37]

$$\left|\Psi_{\mathrm{sp}}^{(\mathrm{s})}\right\rangle \rightarrow \frac{1}{\sqrt{2}}(|S\rangle_1|S\rangle_2 + |C\rangle_1|C\rangle_2), \tag{5}$$

$$\left|\Psi_{\mathrm{sp}}^{(\mathrm{as})}\right\rangle \rightarrow \frac{1}{\sqrt{2}}(|S\rangle_1|C\rangle_2 - |C\rangle_1|S\rangle_2). \tag{6}$$

Here $|C\rangle$ ($|S\rangle$) defines the photon occupying the cosine (sine) standing wave. According to (5) and (6), both photons are always found in the same standing wave if they originate from the symmetric spatial wavefunction, while the photons are split evenly between the sine and cosine waves if they originate from the anti-symmetric spatial wavefunction. Thus, to perform *probabilistic two-photon absorption* and *deterministic one-photon absorption* of the corresponding wavefunctions, the coherent absorber (dissipative beamsplitter) should selectively absorb one of the standing waves while not affecting the other. This non-unitary transformation can be achieved by ultrathin absorptive layer(s), which are transparent if placed at the node(s) of a standing wave while completely dissipating excitations if placed at the anti-node(s). For instance, as shown in Fig. 2C, a single absorptive layer (lossy beamsplitter) may be placed at the anti-node of the cosine standing wave to achieve absorption of its excitations. At the same time, this layer would be located at the node of the sine wave and, therefore, will not interact with it. Next, the second beamsplitter-like transformation brings us back to the travelling waves picture [37], Fig. 2D. Here, two photons from the input symmetric wavefunction that survive the absorption are coherently split evenly between $|L\rangle$ and $|R\rangle$ outgoing waves, restoring the left distribution in Fig. 1B. Similarly, one photon from the input anti-symmetric wavefunction that survives the absorption is coherently split between the outgoing waves, restoring the right distribution in Fig. 1B. Remarkably, such lossy beamsplitter constructed to selectively absorb standing waves of different symmetry would also have the required optical response (4) [37]. The same result can be achieved by replacing a single absorptive layer with multiple layers placed at different anti-nodes of the cosine or sine standing wave. We emphasize that to observe the anti-HOM interference, photon pairs should interfere on the lossy beamsplitter consisting of absorptive layer(s) of the *subwavelength thickness [45-47,50,51]*, as only this design may guarantee the required non-unitary transformation.

*Bosonic and fermionic interference with entangled photons*.-We generate entangled photons via type-II degenerate spontaneous parametric down-conversion (SPDC). Non-linear BBO crystal is pumped by a 405 nm continuous wave laser (Omicron LuxX 405-300, single-mode), generating SPDC photon pairs at a wavelength of $\lambda$=810 nm. The SPDC photons are counter propagated by mirrors (Ms), Fig. 3A. Bell state control is implemented by a set of quarter-half-quarter wave plates [60]. Consequently, the two-photon polarization state, $\frac{1}{\sqrt{2}}\left(|H\rangle_1|V\rangle_2 + e^{i\varphi}|V\rangle_1|H\rangle_2\right)$, can be switched between symmetric, $\varphi = 0$, and anti-symmetric, $\varphi = \pi$, Bell states. We verified the preparation of Bell states by using analyzers with orientation angles $\theta_1$ and $\theta_2$ and single-photon avalanche detectors SPAD-1 and SPAD-2 (Excelitas, SPCM-AQRH-14-FC with an efficiency of 60% at 810 nm) as shown in circular dashed boxes in Fig. 3A. Polarization correlations are measured as a function of $\theta_1$ while $\theta_2$ is fixed at one of the four values: 0, $\pm\pi/4$ or $\pi/2$. As expected, symmetric, Fig. 3B, and anti-symmetric, Fig. 3C, Bell states exhibit identical correlations in horizontal (H) and vertical (V) bases ($\theta_2 = 0$ and $\theta_2 = \pi/2$) and opposite correlations in diagonal (D) and anti-diagonal (A) bases ($\theta_2 = \pm\pi/4$). The Bell parameter $S$ for the symmetric (anti-symmetric) Bell state is 2.63 (2.60) where $\mathrm{S} = \sqrt{2}(\mathrm{V}_{\mathrm{H/V}} + \mathrm{V}_{\mathrm{D/A}})$ and $\mathrm{V}_{\mathrm{H/V}}$ ($\mathrm{V}_{\mathrm{D/A}}$) is the visibility of interference fringes in H/V (D/A) basis [47]. The value of $S$ greater than 2 (capped at 2.828) corresponds to non-classical correlations.

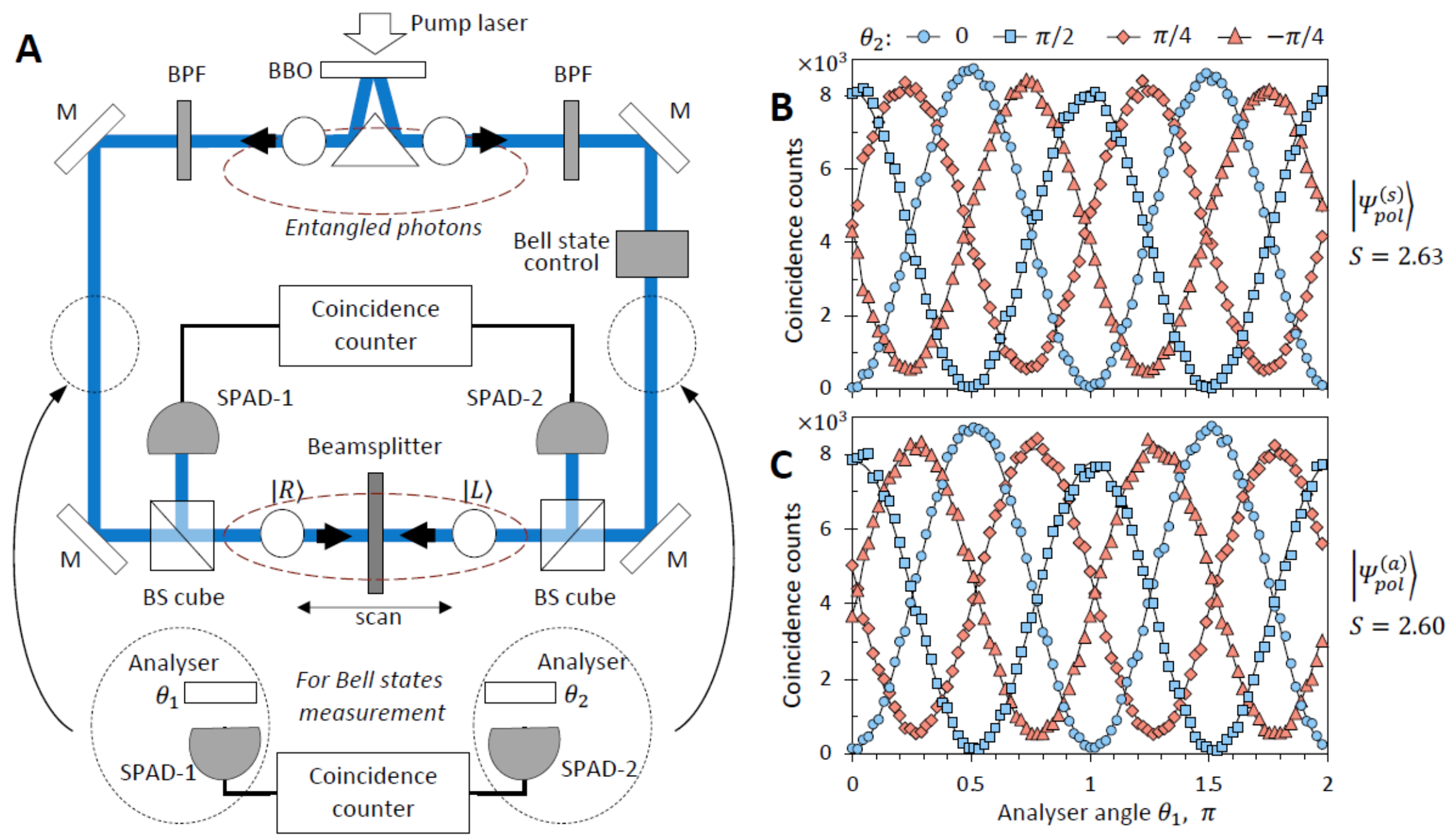

FIG. 3. **Bosonic and fermionic interference with entangled photons**. A: Schematic of the setup for dissipation-free and dissipative interference with entangled photons. Preparation of symmetric (B) and anti-symmetric (C) Bell states is verified through measurement of polarization correlations with apparatus shown in inset in A. All lines in B and C are fitting curves.

Each photon from the entangled pair passes through a bandpass filter (BPF) with the FWHM of 10 nm centered at 810 nm to ensure spectral indistinguishability. Transverse spatial modes are matched by sending photons through single-mode fibers (not shown in Fig. 3A). Additional wave plates are used to compensate for polarization distortion in fibers. Finally, entangled photons interfere on a lossless or lossy beamsplitter placed in the middle of the interferometer. As a lossless beamsplitter, we use a 100 nm thick silicon nitride (SiN) film with reflectance (transmittance) of around 35% (65%) at 810 nm. The counter-propagating geometry in the experiment prevented us from using conventional 50:50 beamsplitters designed for the angular incidence of light. As a lossy beamsplitter (coherent absorber) with $t \approx +r \approx 1/2$, we use two 5 nm thick chromium layers deposited on both sides of the SiN film with a thickness $d = 200$ nm where $d \approx \lambda/(2n_{SiN})$ and refractive index $n_{SiN} \approx 2$. This structure is designed to place the dissipative metallic layers at the anti-nodes (nodes) of the sine (cosine) standing wave. Beamsplitter cubes (BS cubes) direct the outgoing photons to SPAD-1 and SPAD-2, Fig. 3A. By scanning the beamsplitter's position around the center of the interferometer, we vary the degree of photons overlap at the beamsplitter (time of arrival). In the experiment, we measure coincidence counts (IDQ, ID800-TDC) between detectors SPAD-1 and SPAD-2 as a function of the beamsplitter position with the coincidence window of 1.6 ns.

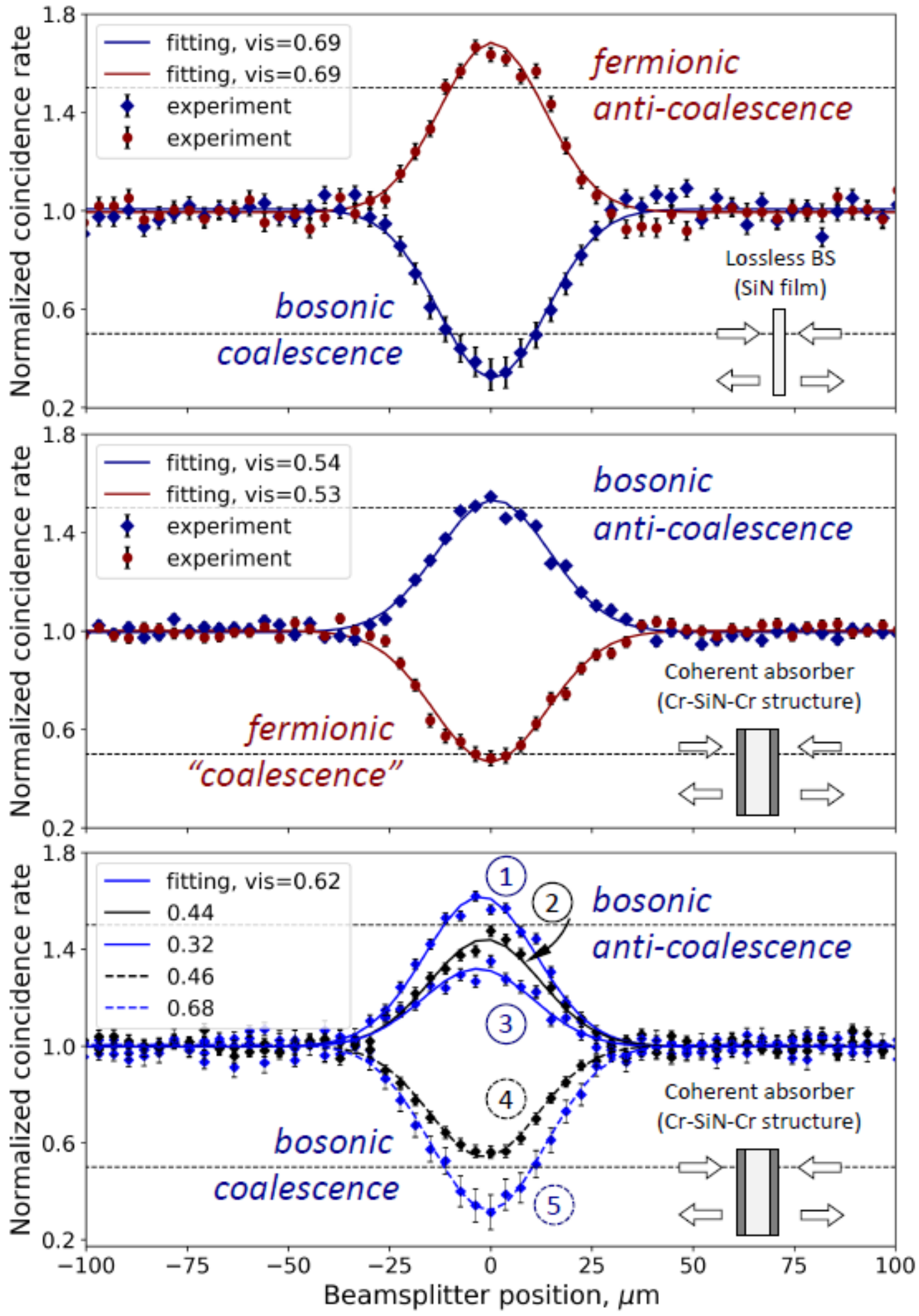

FIG. 4. **Two-photon interference in the absence and presence of dissipation.** A: The HOM experiment with a lossless beamsplitter (inset) and entangled photons. Each point is acquired for 2 seconds, and level 1 corresponds to ~750 coincidence counts. B: The anti-HOM experiment with a coherent absorber, or lossy beamsplitter (SiN spacer with chromium (Cr) layers deposited on both sides). Each point is acquired for 25 seconds, and level 1 corresponds to ~2150 coincidence counts. C: Control of interference of non-entangled photons by changing parameters of a coherent absorber (lossy beamsplitter). All lines are Gaussian fitting, visibilities are shown in the boxes. Dashed horizontal lines show corresponding classical limits. Error bars are due to the shot noise of the counting. Insets show schematics of the corresponding samples used in the experiment.

First, we illuminate the lossless beamsplitter by photons with the symmetric spatial wavefunction, $|\Psi^{(s)}_{\mathrm{pol}}\rangle \otimes |\Psi^{(s)}_{\mathrm{sp}}\rangle$ (blue diamonds in Fig. 4A). When photons from the pair arrive at the

beamsplitter at different times ('Beamsplitter position' $\lesssim -30\mu m$ and $\gtrsim 30\mu m$ ), interference does not happen, and there is a close to 0.5 probability that photons take different output ports. These events generate the 'reference' level of coincidence counts. When photons overlap at the beamsplitter, the HOM effect occurs: photons coalesce with the characteristic dip in coincidence counts. Next, the lossless beamsplitter is illuminated by photons with the anti-symmetric spatial wavefunction, $|\Psi_{\mathrm{pol}}^{(as)}\rangle \otimes |\Psi_{\mathrm{sp}}^{(\mathrm{as})}\rangle$ (red circles in Fig. 4A). In the absence of interference, around half of photon pairs take the same output port. These events are suppressed when photons are overlapped at the beamsplitter, and photons tend to leave the beamsplitter through different output ports. The number of coincidence counts rises, and the characteristic peak is observed. We note that the 'bosonic' dip does not reach the minimum value of 0, and the 'fermionic' peak does not reach the maximum value of 2 in Fig. 4A due to unequal reflectance and transmittance of the beamsplitter and imperfect Bell states (see Supplementary Material).

When the lossless beamsplitter is replaced by the coherent absorber (lossy beamsplitter), the result of interference changes drastically, Fig. 4B. For symmetric spatial wavefunction, $|\Psi_{\mathrm{pol}}^{(s)}\rangle \otimes \left|\Psi_{\mathrm{sp}}^{(\mathrm{s})}\right\rangle$ (blue diamonds in Fig. 4B), in the absence of interference, there is a 0.5 chance that one photon will be absorbed and the survived photon does not contribute to coincidence counts. The probability of the coincidence count event is $2 * (0.25 * 0.25) = 0.125$ (both photons are transmitted with the probability of $0.25 * 0.25$ and both photons are reflected with the same probability). When photons are overlapped at the beamsplitter, single-photon absorption is suppressed, at the same time increasing the probability of photons taking different output ports. As a result, a peak in coincidence counts is observed. For anti-symmetric spatial wavefunction, $|\Psi_{\mathrm{pol}}^{(as)}\rangle \otimes \left|\Psi_{\mathrm{sp}}^{(\mathrm{as})}\right\rangle$ (red circles in Fig. 4B), the probability of two photons to be detected at different output ports is, again, 0.125, when photons arrive at different time. When photons are overlapped at the beamsplitter, deterministic one-photon absorption takes place. Since the survived photon does not produce coincidence counts, the dip in coincidence counts is detected. We note that the precise tuning of the beamsplitter's parameters and exploitation of the maximally entangled Bell states would allow reaching the maximum possible level of 2 for the 'bosonic' peak and the minimum possible level of 0 for the 'fermionic' dip (see Supplementary Material). Importantly, coincidence counts cross the classical limits of 0.5 or 1.5 indicating the quantum regime of interference [22,61]. To eliminate the contribution of non-ideal entanglement in results in Fig. 4B, we repeat the experiment by placing identically oriented polarizers in each arm of the interferometer (at dashed circles in Fig. 3A). As photons have the same polarization, only symmetric spatial wavefunction (1) is allowed. Line 1 in Fig. 4C shows the result of this experiment, in which the 'bosonic' peak rises ~10% higher compared to Fig. 4B.

As discussed above, the anti-coalescent component of two-photon interference is defined by the coefficient $P_{11} = |t^2 + r^2|^2$. By exploiting lossy beamsplitters (coherent absorbers), one may access a broad range of values for $t$ and $r$ by adjusting both their absolute values and mutual phase, thus changing the output two-photon statistics (see Supplementary Material). For the chromium-SiN-chromium structure of lossy beamsplitter, the thickness of the chromium layers defines the absolute values of $t$ and $r$, not affecting their mutual phase (a few nanometers thick layers introduce negligible phase shifts). On the other hand, the thickness of the SiN layer simultaneously affects both the magnitudes of $t$ and $r$ and their mutual phase (see Fig. 5 in Ref. [62]). We demonstrate control over two-photon statistics of the bosonic spatial wavefunction of non-entangled photons for different thicknesses of chromium and SiN layers. Lines 1-3 in Fig. 4C show results of the two-photon interference with absorbers fabricated by deposition of absorptive chromium layers of 5 nm (line 1), 4.5 nm (line 2) and 4 nm (line 3) thick on both sides of SiN substrate with $d = 200$ nm. The magnitude of the anti-coalescent peak, $P_{11}$, decreases as losses of the beamsplitter are decreased. In this case,

the absolute value of $t$ and $r$ changes while their mutual phase is fixed. To change the phase relation between $t$ and $r$, we change the SiN substrate thickness. Lines 4 and 5 in Fig. 4C show results of the interference on chromium layers of 5 nm (line 4) and 4 nm (line 5) deposited on both sides of SiN substrate with $d = 1000$ nm. The dip in coincidences indicates that the phase relation between $t$ and $r$ is close to that of a lossless beamsplitter. Similarly, a change in the absolute value of $t$ and $r$ allows altering the amplitude of the dip or the degree of photon coalescence. The output statistics can be changed dynamically if the optical thickness of a substrate is controlled in situ, for instance, by using a liquid crystal.

*Discussions and conclusion*.**-**We experimentally demonstrated the anti-HOM interference of symmetric and anti-symmetric spatial wavefunctions of entangled photons assisted by coherent perfect absorption. We showed that bosonic spatial wavefunctions, widely accepted as exhibiting coalescent behavior, may display the anti-coalescent regimes of interference, while fermionic wavefunctions, known as exhibiting anti-coalescent behavior, may reveal coalescent-like regimes of interference. The anti-HOM interference is a fundamental effect relevant to the interference of any two particles in the presence of coherent absorption and may be used to substitute the HOM interference in practical applications. Moreover, the ability to switch between the fermionic behavior of bosons and the bosonic behavior of fermions may enrich quantum information and metrology protocols based on entangled photons, and can be of interest for quantum simulations [63].

Here, we achieved control over two-photon statistics via the interference of quantum light on a thin coherent absorber. This method can be further implemented in integrated platforms such as optical waveguides and fiber networks [50,51]. We note that the functionality of the coherent perfect absorber can be realized in a cascaded lossless beamsplitters design, as discussed in Ref. [37], although it comes at the expense of setup complications or an increased footprint for integrated implementation.

Beamsplitter-like transformation appears in the interaction between quantum light and matter [64] where light absorption happens due to the excitation of inner states of matter. From this point of view, the anti-HOM interference could provide novel approaches for protocols of quantum memory [65-68], quantum networks, and quantum communication [69,70]. Interference of photons in the presence of dissipation channels may also be extended to other schemes of coherent light illumination, including virtual absorption [71], absorption in coupled matter-cavity systems [72-74], and non-linear regimes of coherent absorption [75].

Importantly, demonstrated anti-HOM interference may be considered as a manifestation of the fundamental underlying process – redistribution of photons originally contained in propagating waves between standing waves. This picture would allow consideration of coherent phenomena in quantum optics under a new perspective, and interaction protocols based on this mechanism may be realized.

The data that supports the findings of this study are openly available in NTU research data repository DR-NTU (Data) at (*to be provided*).

**Acknowledgment**

This work was supported by the Singapore Ministry of Education [Grant No. MOE2016-T3-1-006 (S)], the Quantum Engineering Programme of the Singapore National Research Foundation (QEP-P1 and NRF-QEP2-01-P01), and the UK's Engineering and Physical Sciences Research Council (Grant No. EP/M009122/1).